\documentclass[twocolumn,aps,prl,preprintnumbers]{revtex4}
\usepackage{bbm}
\usepackage[latin9]{inputenc}
\usepackage{amsmath}
\usepackage{amssymb}
\usepackage{graphicx}
\usepackage{mathrsfs}
\usepackage{amsfonts}
\usepackage{amsthm}
\usepackage{color}
\usepackage{txfonts}
\usepackage{tabularx}
\usepackage[colorlinks=true,citecolor=blue,linkcolor=blue,urlcolor=blue,anchorcolor=blue]{hyperref}%
\hypersetup{colorlinks=true,citecolor=blue,linkcolor=blue,urlcolor=blue}
\providecommand{\U}[1]{\protect\rule{.1in}{.1in}}
\makeatletter
\@ifundefined{textcolor}{}
{
\definecolor{BLACK}{gray}{0}
\definecolor{WHITE}{gray}{1}
\definecolor{RED}{rgb}{1,0,0}
\definecolor{GREEN}{rgb}{0,1,0}
\definecolor{BLUE}{rgb}{0,0,1}
\definecolor{CYAN}{cmyk}{1,0,0,0}
\definecolor{MAGENTA}{cmyk}{0,1,0,0}
\definecolor{YELLOW}{cmyk}{0,0,1,0}
}
\makeatother
\begin{document}
\title{Negative Gilbert damping in cavity optomagnonics}
\author{Yunshan Cao}
\email[Corresponding author: ]{yunshan.cao@uestc.edu.cn}
\author{Peng Yan}
\email[Corresponding author: ]{yan@uestc.edu.cn}
\affiliation{School of Electronic Science and Engineering and State Key Laboratory of Electronic Thin Films and Integrated Devices, University of
Electronic Science and Technology of China, Chengdu 610054, China}

\begin{abstract}
Exceptional point (EP) associated with the parity-time ($\mathcal{PT}$) symmetry breaking is receiving considerable recent attention by the broad physics community. By introducing balanced gain and loss, it has been realized in photonic, acoustic, and electronic structures. However, the observation of magnonic EP remains elusive. The major challenge is to experimentally generate the negative Gilbert damping, which was thought to be highly unlikely but is demanded by the $\mathcal{PT}$ symmetry. In this work, we study the magneto-optical interaction of circularly-polarized lasers with a submicron magnet placed in an optical cavity. We show that the off-resonant coupling between the driving laser and cavity photon in the far-blue detuning can induce the magnetic gain (or negative damping) exactly of the Gilbert type. A hyperbolic-tangent function ansatz is found to well describe the time-resolved spin switching as the intrinsic magnetization dissipation is overcome. When the optically pumped magnet interacts with a purely lossy one, we observe a phase transition from the imbalanced to passive $\mathcal{PT}$ symmetries by varying the detuning coefficient. Our findings provide a feasible way to manipulate the sign of the magnetic damping parameter and to realize the EP in cavity optomagnonics.
\end{abstract}
\maketitle

\emph{Introduction.}---One of the most fundamental principles in quantum mechanics is that a physical observable should be described by a Hermitian operator to guarantee real eigenvalues \cite{Dirac1958}. However, Bender and Boettcher \cite{Bender1998} reported a class of non-Hermitian Hamiltonians that allow entirely real spectrum as long as the combined parity ($\mathcal{P}$) and time ($\mathcal{T}$)-reversal symmetries are respected. By tuning system parameters, both the eigenvalues and eigenstates of the $\mathcal{PT}$-symmetric Hamiltonian simultaneously coalesce \cite{Bender2002,Bender2002a}, giving rise to a non-Hermitian degeneracy called exceptional point (EP). The nature around the EP that is accompanied by a phase transition can trigger many intriguing phenomena,
such as unidirectional invisibility \cite{Feng2012,Peng2014},
loss-induced laser suppression and revival \cite{Peng2014b} and optical transparency \cite{Guo2009},
laser mode selection \cite{Feng2014},
and EP enhanced sensing \cite{Chen2017,Hodaei2017,Chen2018,Zhong2019}.
Over the past decades, the experimental observation of EPs has been realized in a broad field of photonics \cite{Konotop2016,Wen2018,El-Ganainy2018,Oezdemir2019}, acoustics \cite{Zhu2014,Jing2014}, and electronics \cite{Schindler2011,Schindler2012,Quijandria2018}.
Very recently, the concept of $\mathcal{PT}$ symmetry is attracting significant attention in spintronics and magnonics \cite{Lee2015,Yang2018,Harder2017,Galda2016,Galda2018,Zhang2017,Cao2019,Yuan2020,Yu2020}. The simplest way to obtain a $\mathcal{PT}$-symmetric system consists in coupling two identical subsystems, one with gain and the other with equal amount of loss. The composite system is $\mathcal{PT}$ symmetric because space reflection interchanges the subsystems, and time reversal interchanges gain and loss. Indeed, a $\mathcal{PT}$-symmetric magnetic structure composed of two identical ferromagnets with balanced gain and loss was first proposed by Lee \emph{et al.} \cite{Lee2015} and subsequently investigated by Yang \emph{et al}. \cite{Yang2018}. One recent breakthrough was made by Liu \emph{et al}. \cite{Liu2019} who reported EP in passive $\mathcal{PT}$-symmetric devices in the form of a trilayer structure with two magnetic layers of different (positive) Gilbert damping. However, the experimental observation of genuine $\mathcal{PT}$ symmetry for magnons (the quanta of spin waves)---as elementary excitations in ordered magnets---is still elusive. The difficulty lies in that the Gilbert damping can hardly be tuned to be negative \cite{Gilbert2004,Hickey2009}.

The past ten years have witnessed the development and application of spin cavitronics, allowing cavity photons resonantly coupled to magnons with the same microwave frequency \cite{Huebl2013,Bhoi2014,Tabuchi2014,Zhang2014,Goryachev2014,Zhang2015,
Bai2015,Cao2015,Flaig2016,Rameshti2015,Tabuchi2015,Quirion2017}. One recent trend beyond microwaves is the realization of the parametric coupling between optical lasers and magnons, that would generate interesting new opportunities. Tantalizing physics indeed has been demonstrated, such as nonreciprocal Brillouin light scattering \cite{Osada2018}, microwave-to-optical converting \cite{Andrews2014,Zhang2016}, optical cooling of magnons \cite{Sharma2018}, etc. In these studies, considerable interests have been drawn to the scalar properties of magnons, e.g., magnon number (population), temperature, and chemical potential, which is successful to describe the small-angle spin precession. In contrast, their vectorial behavior, i.e., the full time-evolution of the magnetic moment driven by optical lasers, remains largely unexplored, with few exceptions \cite{Kusminskiy2016}. It has been shown that a ferromagnetic-to-antiferromangetic phase transition may emerge in the vicinity of the magnonic EP \cite{Yang2018}. In such case, the magnetic moment would significantly deviate from its equilibrium direction, and a vectorial field description becomes more relevant than a scalar one.

\begin{figure}[!htbp]
  \centering
  \includegraphics[width=\linewidth]{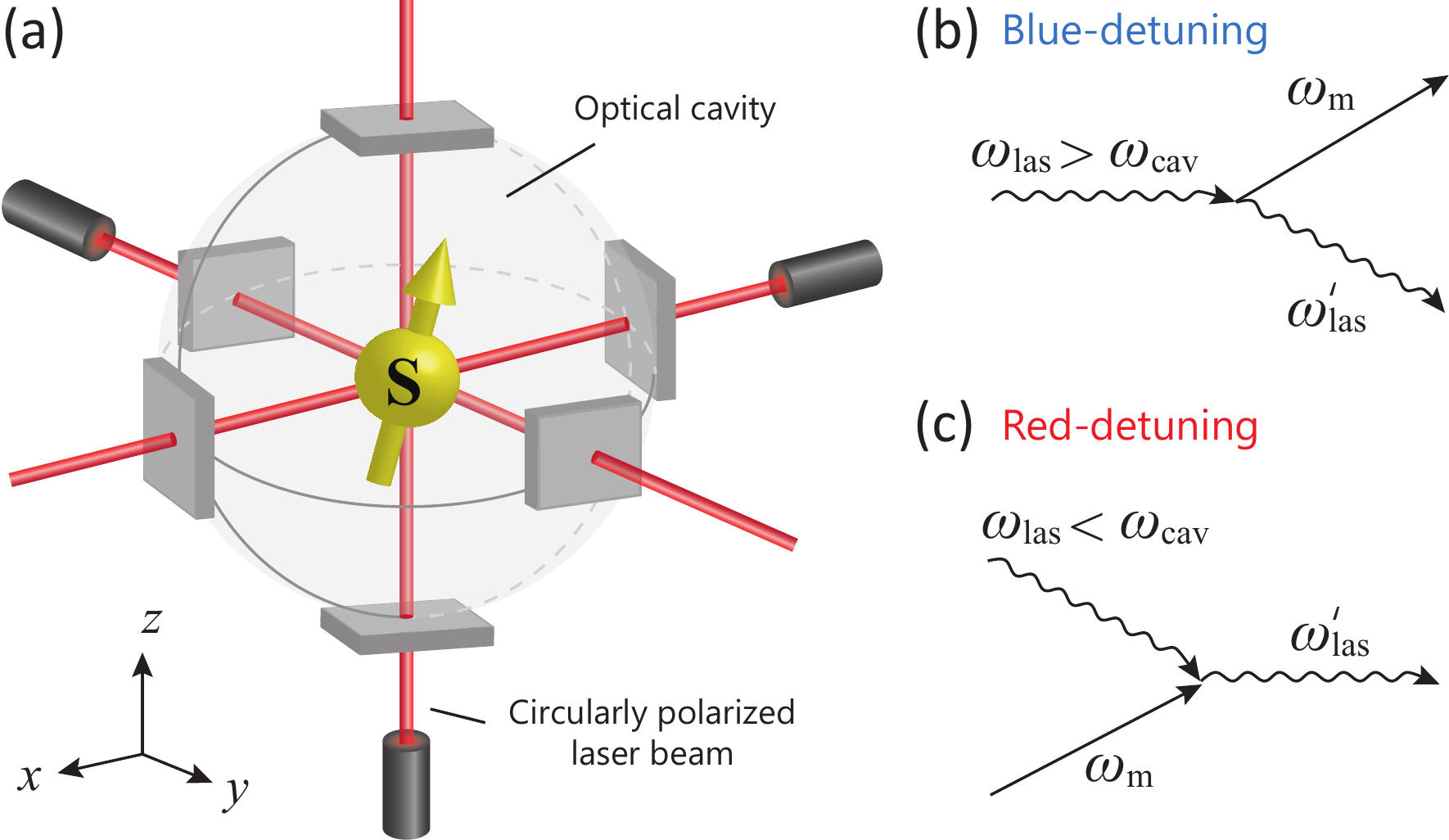}
  \caption{(a) Schematic illustration of a macrospin $\textbf{S}$ interacting with three orthogonally propagating circularly-polarized lasers (red beams) in an optical cavity. Off-resonant coupling between the driving laser ($\omega_{\text{las}}$) and the cavity photon ($\omega_{\text{cav}}$) mediated by magnons ($\omega_{\text{m}}\ll\omega_{\text{cav}}$) in the blue (b) and red (c) detuning regimes.}\label{system}
\end{figure}

In this Letter, we propose to realize the negative Gilbert damping by considering the optomagnonic interaction between three orthogonally propagating circularly-polarized lasers and a submicron magnet placed in an optical cavity [see Fig. \ref{system}(a)]. By solving the coupled equations of motion and integrating the photon's degree of freedom, we derive the analytical formula of the optical torque acting on the macrospin. In the far-blue detuning, we find that the optical torque exactly takes the Gilbert form $-\frac{\alpha_\mathrm{opt}}{S}\dot{\mathbf{S}}\times\mathbf{S}$ with $\alpha_\mathrm{opt}>0$ (see below). The total Gilbert damping becomes negative when the intrinsic dissipation is overcome. In such case, a hyperbolic-tangent function ansatz is found to well describe the time-resolved spin switching. We further study the optically pumped spin interacting with a purely lossy one, and observe a phase transition from the imbalanced to passive $\mathcal{PT}$ symmetries by varying the detuning parameter.

\emph{Model.}---The proposed setup is schematically plotted in Fig. \ref{system}(a). Three circularly-polarized laser beams propagating respectively along $x, y, z$ directions drive the parametric coupling with a macrospin $\mathbf{S}=(\hat{S}_{\!x},\hat{S}_{\!y},\hat{S}_{\!z})$ inside the optical cavity. The Hamiltonian reads
\begin{equation}\label{ham}
    \mathcal{H}= -\hbar\omega_0\hat{S}_z
    -\hbar\sum_{j=x,y,z}\left(\Delta_j-g_j\hat{S}_j\right)\hat{c}^\dagger_j \hat{c}_j
    +\mathcal{H}_\mathrm{dr},
\end{equation}
where $\omega_0=\gamma B_0$ is the Larmor frequency around the external magnetic field $\mathbf{B}_0$ pointing to the negative $z$-direction with $\gamma$ being the gyromagnetic ratio, $\Delta_j=\omega_{\mathrm{las},j}-\omega_\mathrm{cav}$ is the detuning between the laser frequency $\omega_{\mathrm{las},j}$ and the cavity resonant frequency $\omega_\mathrm{cav}$, and $\hat{c}_j^\dagger (\hat{c}_j)$ is the creation (annihilation) operator of the optical cavity photons, with $j=x,y,z$. The coupling strength $g_j$ between the spin and optical photon originates from the Faraday-induced modification of the electromagnetic energy in ferromagnets \cite{Landau1984}. The last term describes the interaction between the driving laser and the cavity photon $\mathcal{H}_\mathrm{dr}=i\hbar\sum_j (A_j\hat{c}^\dagger_j-\mathrm{h.c.})$,
where $A_j=(2\kappa_j P_{j}/\hbar\omega_{\mathrm{las},j})^{1/2}$ is the field amplitude, with $\kappa_j$ the laser loss rate and $P_{j}$ being the driving power.

The Heisenberg-Langevin equations of motion for coupled photons and spins are expressed as ($o\equiv \langle \hat{o}\rangle$),
\begin{subequations}
  \begin{eqnarray}\label{fastcavity}
  \dot{c}_j &=& (i\Delta_j-\kappa_j) c_j-ig_jS_j c_j+ A_j, \\
  \label{spindynamic}
  \dot{S}_x &=& \omega_0 S_y+g_y n_y S_z-g_z n_z S_y, \\
  \dot{S}_y &=& -\omega_0 S_x-g_x n_x S_z+g_z n_z S_x,\\
   \label{spindynamic2}
  \dot{S}_z &=& -g_y n_y S_x+g_x n_x S_y,
\end{eqnarray}
\end{subequations}
where $n_j=\langle \hat{c}^\dagger_j\hat{c}_j \rangle$ is the average photon number in the cavity. Because the spin dynamics usually is much slower than optical photons, one can expand the cavity photon operator as $c_j(t)\approx c_{j0}(t)+c_{j1}(t)+\cdots$, in orders of $\dot{S}_j$. Equation (\ref{fastcavity}) then can be recast in series
\begin{subequations}
  \begin{eqnarray}
  0 &=& (i\Delta_j-\kappa_j) c_{j0}-ig_jS_j c_{j0}+ A_j, \\
  \dot{c}_{j0} &=&  (i\Delta_j-\kappa_j) (c_{j0}+c_{j1})-ig_jS_j (c_{j0}+c_{j1})+ A_j,
\end{eqnarray}
\end{subequations}
by keeping up to the first-order terms. We can therefore derive the formula of photon number in the cavity
\begin{eqnarray}\label{photonno}
 n_j(t) &\approx& |c_{j0}|^2+2\mathrm{Re}[c^*_{j0}c_{j1}]\nonumber\\
 &=&\frac{A_j^2}{(\Delta_j-g_j S_j)^2+\kappa_j^2}-\frac{4\kappa_j A_j^2g_j(\Delta_j-g_j S_j)}{\left[(\Delta_j-g_j S_j)^2+\kappa_j^2\right]^3}\dot{S}_j.
\end{eqnarray}
Substituting (\ref{photonno}) into Eqs. (\ref{spindynamic})-(\ref{spindynamic2}), we obtain
\begin{equation}\label{Seq1}
  \dot{\mathbf{S}}=-\gamma\mathbf{S}\times\mathbf{B}_\mathrm{eff}
  +\frac{\alpha}{S}(\dot{\mathbf{S}}\times\mathbf{S})-\mbox{\boldmath$\beta$}_\mathrm{opt}\times\mathbf{S},
\end{equation}
where the effective magnetic field $\mathbf{B}_\mathrm{eff}=-B_0\mathbf{e}_z+\mathbf{B}_\mathrm{opt}$ includes both the external magnetic field and the optically induced magnetic field
\begin{equation}\label{bopt}
   \mathbf{B}_\mathrm{opt}=\sum_{j}\frac{\gamma^{-1}g_jA_j^2}{(\Delta_j-g_j S_j)^2+\kappa_j^2}\mathbf{e}_j,
  \end{equation}
which is the zeroth-order of $\dot{S}_j$. The second term in the right hand side of (\ref{Seq1}) is the intrinsic Gilbert damping torque, with $S=|\mathbf{S}|$ the total spin number and $\alpha>0$ being the intrinsic Gilbert damping constant. The last term in (\ref{Seq1}) represents the optical torque with the anisotropic effective field
\begin{equation}\label{betaopt}
   \mbox{\boldmath$\beta$}_\mathrm{opt}=\sum_j\frac{4\kappa_j A_j^2g^2_j(\Delta_j-g_j S_j)}{\left[(\Delta_j-g_j S_j)^2+\kappa_j^2\right]^3}\dot{S}_j\mathbf{e}_j,
\end{equation}
which is linear with the first-order time-derivative of $S_j$. Below, we show that the anisotropic nature of (\ref{betaopt}) can be smeared out under proper conditions.

\emph{Negative Gilbert damping.}---To obtain the optical torque of exactly the Gilbert form, we make two assumptions: (i) the three laser beams are identical, i.e., $A_j=A, g_j=g,\kappa_j=\kappa,$ and $\Delta_j=\Delta$; (ii) the optomagnonic coupling works in the far detuning regime, i.e., $|\eta|\gg 1$ with $\eta=\Delta/(gS$), which allows us to drop the $g_jS_j$ terms in Eq. (\ref{betaopt}). The optically induced effective fields then take the simple form
\begin{equation}
   \mathbf{B}_\mathrm{opt}=\frac{\gamma^{-1}gA^2}{\Delta^2+\kappa^2}\sum_{j}\mathbf{e}_j,
\end{equation}
and
\begin{equation}\label{aopt}
 \mbox{\boldmath$\beta$}_\mathrm{opt}=\frac{\alpha_\mathrm{opt}}{S}\dot{\mathbf{S}},~~\text{with}~~
 \alpha_\mathrm{opt}=\frac{4\kappa A^2g^2S\Delta}{(\Delta^2+\kappa^2)^3}
\end{equation}
being the laser-induced magnetic gain or loss that depends the sign of the detuning $\Delta$. Based on the above results, we finally obtain the optically modulated spin dynamics
\begin{equation}\label{Seq2}
  \dot{\mathbf{S}}=-\gamma\mathbf{S}\times\mathbf{B}_\mathrm{eff}
  +\frac{\alpha_\mathrm{eff}}{S}(\dot{\mathbf{S}}\times\mathbf{S}),
\end{equation}
with $\alpha_\mathrm{eff}=-\alpha_\mathrm{opt}+\alpha$. One can observe that a negative effective Gilbert constant ($\alpha_\mathrm{eff}<0$) emerges in the far-blue detuning regime, i.e., $1<\eta<\eta_\mathrm{c}$. In case of the red detuning ($\eta<0$), we have $\alpha_\mathrm{opt}<0$, which indicates the enhancement of the magnetic attenuation. In the deep-blue detuning regime ($\eta>\eta_\mathrm{c}$), driving lasers can still generate the magnetic gain ($\alpha_\mathrm{opt}>0$) but cannot compensate the intrinsic dissipation, i.e., $0<\alpha_\mathrm{opt}<\alpha$. Here $\eta_\mathrm{c}$ is the critical detuning parameter at which the effective Gilbert damping vanishes. The physics can be understood from the diagram plotted in Figs. \ref{system}(b) and \ref{system}(c): In the blue detuning regime $(\omega_{\text{las}}>\omega_{\text{cav}})$, microwave magnons are emitted in the non-resonant interaction between the driving laser and the cavity photon, representing a magnetic gain. On the contrary, they are absorbed in the red detuning $(\omega_{\text{las}}<\omega_{\text{cav}})$, manifesting a magnon absorption or cooling. Below we discuss practical materials and parameters to realize this proposal.

\begin{figure}
  \centering
  \includegraphics[width=\linewidth]{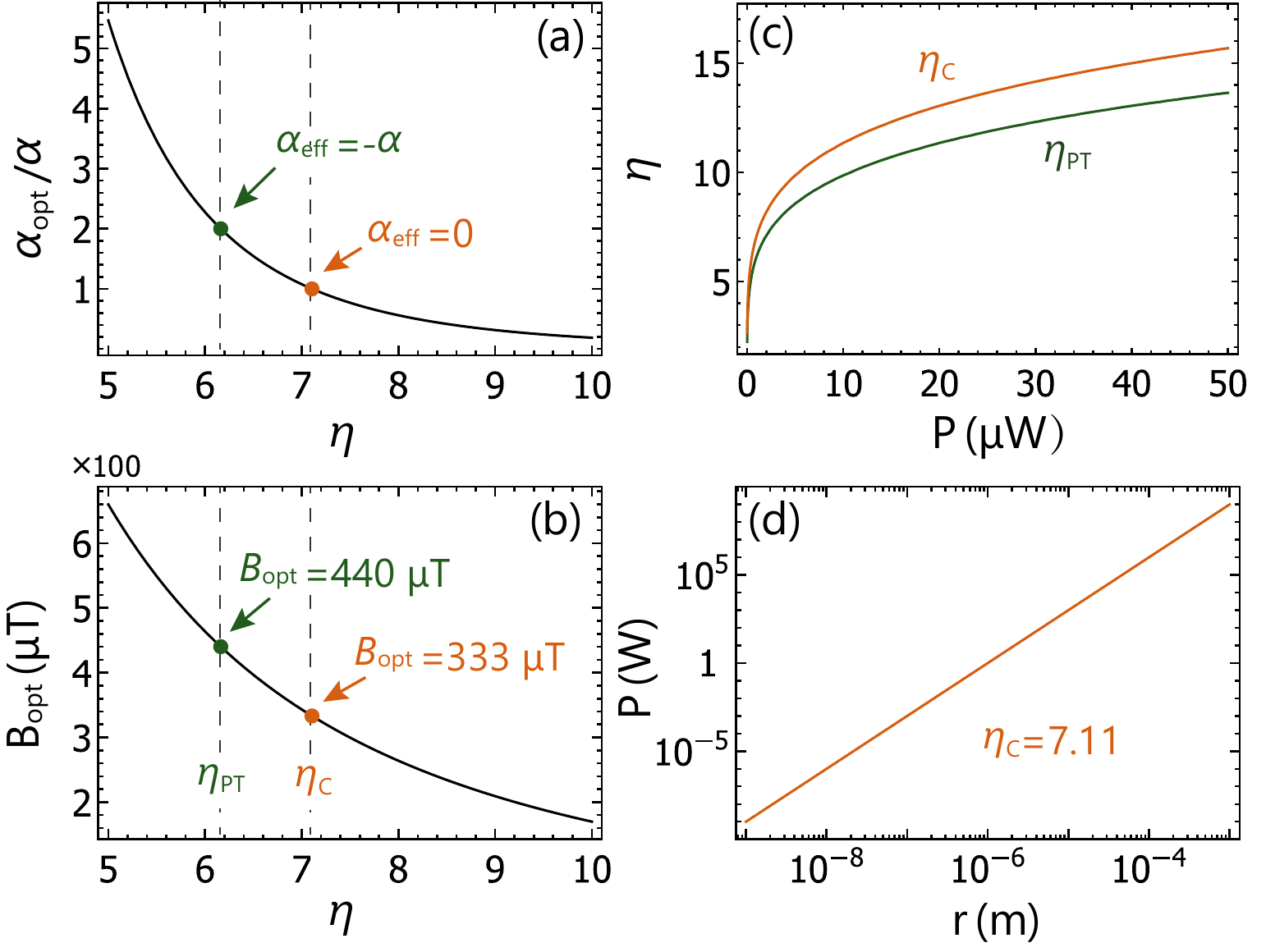}
  \caption{Optically induced magnetic gain (a) and magnetic field (b) vs. the optical detuning parameter $\eta$.
  (c) $\eta_\text{\tiny PT}$ (orange) and $\eta_\text{\tiny C}$ (green) as a function of the driving laser power.
  (d) Radius dependence of the laser power at the compensation point $\eta_\text{\tiny C}=7.11$. }\label{aeff}
\end{figure}

\emph{Materials realizations.}---For a ferromagnetic insulator like yttrium ion garnet (YIG), the intrinsic Gilbert constant $\alpha$ typically ranges $10^{-3}\sim10^{-5}$ \cite{Kajiwara2010,Heinrich2011,Kurebayashi2011}. We take $\alpha=10^{-4}$ in the following calculations. The magneto-optical coupling strength is determined by the Faraday rotation coefficient $\theta_F$ of the materials $gS\simeq c\theta_F/\sqrt{\epsilon_r}$, with $c$ the speed of light and $\epsilon_r$ the relative permittivity (for YIG, we choose $\epsilon_r = 15$ \cite{Sadhana2009} and $\theta_F=188^\circ/ \mathrm{cm}$ \cite{Cooper1968}). We thus have $gS/2\pi\approx 1$ GHz. The optical cavity is set at the resonant frequency $\omega_\mathrm{cav}/2\pi=100$ THz with the loss rate $\kappa/2\pi=1$ GHz. For a YIG sphere of radius $r=10$ nm and spin density $\rho_s\approx 10^{28}$ $\mathrm{m}^{-3}$, we estimate the total spin number $S=\rho_sr^3 \approx 10^4$ and the coupling strength $g/2\pi\approx 0.1$ MHz. Materials parameters are summarized in Table \ref{parameters}. Because $g\ll\kappa$, all interesting physics occurs in the weak coupling regime. A negative $\alpha_{\text{eff}}$ is demanded for realizing the $\mathcal{PT}$ symmetry in magnetic system. Considering the driving laser with a fixed power $P=1$ $\mu$W, \vspace{-1em}
\begin{table}[!htbp]
\centering
\caption{Parameters for optical cavity and YIG.}
\renewcommand\arraystretch{1.3}
\begin{tabular}{>{\centering\arraybackslash}p{0.16\linewidth}>{\centering\arraybackslash}p{0.16\linewidth}
>{\centering\arraybackslash}p{0.16\linewidth}>{\centering\arraybackslash}p{0.16\linewidth}
>{\centering\arraybackslash}p{0.13\linewidth}>{\centering\arraybackslash}p{0.13\linewidth}}
  \hline
  \hline
  $~\omega_\text{cav}/2\pi~$&$~\kappa/2\pi~$ & $\omega_0/2\pi$ & $gS/2\pi$ & $r$ & $\alpha$ \\
  \hline
  100 THz & 1 GHz & 10 GHz & 1 GHz & 10 nm & $10^{-4}$ \\
  \hline
  \hline
\end{tabular}\label{parameters}
\end{table}the effective Gilbert-type magnetic gain is $\alpha_\mathrm{eff}=-\alpha$ at $\eta_\text{\tiny PT}\simeq 6.16$, and the critical gain-loss point $\alpha_\mathrm{eff}=0$ occurs at $\eta_\text{\tiny C}\simeq7.11$, indeed satisfying the large-detuning condition $|\eta|\gg 1$ in deriving (\ref{aopt}). Figure \ref{aeff}(a) shows the monotonically decreasing dependence of the optically induced magnetic gain $\alpha_\mathrm{opt}$ on the detuning parameter $\eta$. The $\eta$-dependence of the optical field is plotted in Fig. \ref{aeff}(b), showing that it monotonically decreases with the increasing of the detuning, too. Enhancing the laser power will push the two critical points $\eta_\text{\tiny C}$ and $\eta_\text{\tiny PT}$ into the deep detuning region, as demonstrated in Fig. \ref{aeff}(c). For a magnetic sphere of larger volume $(1\ \mathrm{\mu m})^3 \sim(1\ \mathrm{mm})^3$ that contains a total spin number $S= 10^{10} \sim 10^{19}$ with the reduced magneto-optical coupling strength $g/2\pi= 10^{-1} \sim10^{-10}$ Hz, the required laser power then should be $6 \sim 15$ orders of magnitude higher than the nm-scale sphere case, as shown in Fig. \ref{aeff}(d).

\begin{figure}[!htbp]
  \centering
  \includegraphics[width=\linewidth]{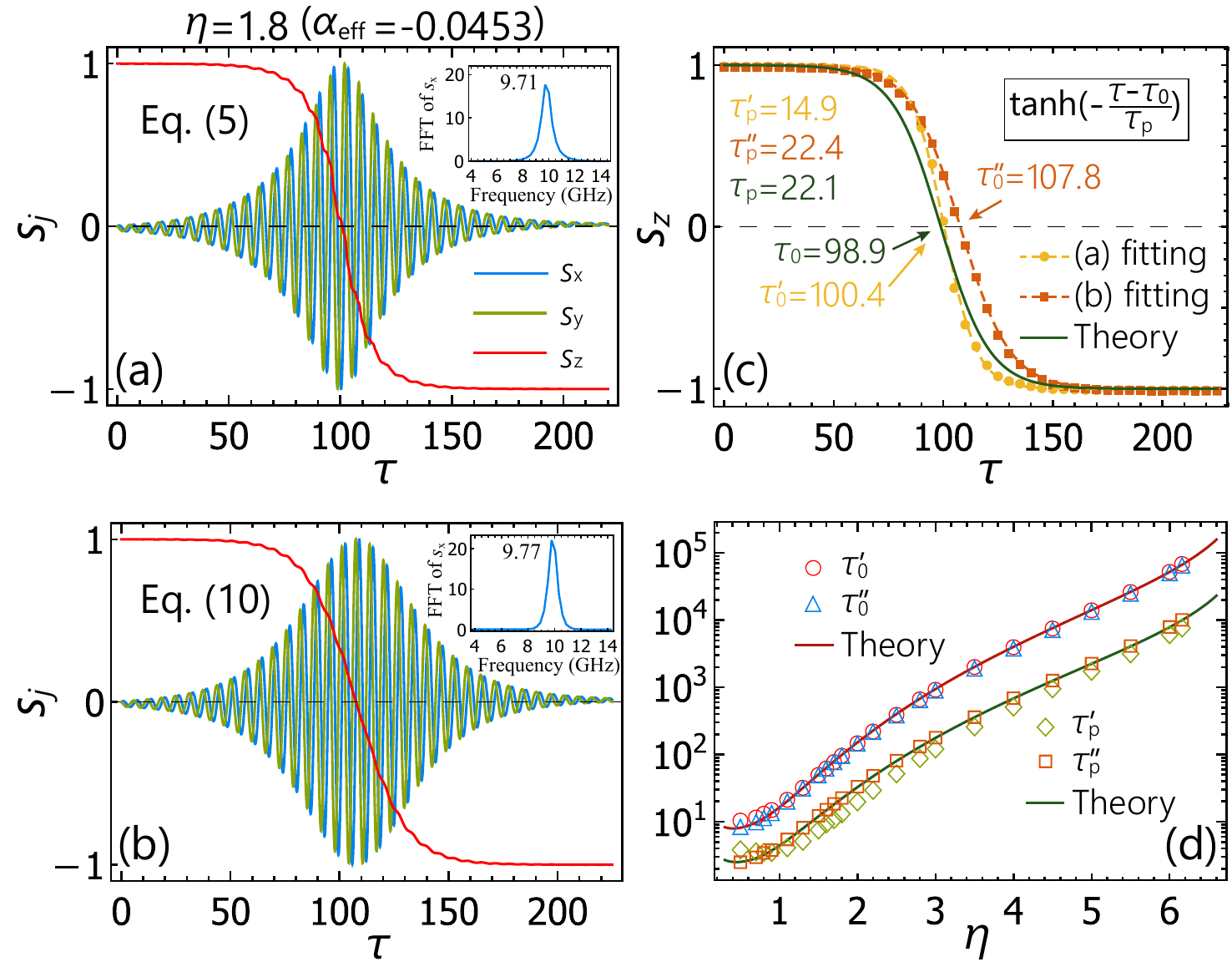}
  \caption{Time evolution of unit spin components $(s_x, s_y, s_z)$ at detuning $\eta=1.8$ based on Eq. (\ref{Seq1}) (a) and Eq. (\ref{Seq2}) (b). Insets show the FFT spectrum of $s_x$. (c) Theoretical fittings of $s_z$ using the hyperbolic-tangent ansatz (\ref{tanheq}) (dashed curves). The solid green curve is the analytical formula without any fitting.
  (d) Numerical results of the $\eta$-dependence of the two characteristic times $\tau_0$ and $\tau_p$, comparing with formula (\ref{taueq}) (solid curves).
    }\label{timesolution}
\end{figure}

\emph{Time-resolved spin flipping.}---To justify the approximation adopted in deriving the Gilbert-type magnetic gain, we directly simulate the time evolution of the unit spin components ($s_j\equiv S_j/S$) based on both Eq. (\ref{Seq1}) and Eq. (\ref{Seq2}). Numerical results are, respectively, plotted in Figs. \ref{timesolution}(a) and \ref{timesolution}(b) for the same detuning parameter $\eta=1.8$ (corresponding to an effective magnetic gain $\alpha_\mathrm{eff}=-0.0453$) and $\omega_0/2\pi=10$ GHz. Both figures show that the very presence of the negative Gilbert damping can flip the spin in a precessional manner, with similar switching curves. The fast Fourier transformation (FFT) analysis of the spatiotemporal oscillation of $s_x$ also confirms this point (see the insets). Although the analytical form of $s_z(t)$ by solving (\ref{Seq1}) generally is unknown \cite{Kikuchi1956,ZZSun2005}, we find an ansatz that can well describe the time-resolved spin switching
\begin{equation}\label{tanheq}
   s_z(\tau)\simeq\tanh\left(-\frac{\tau-\tau_0}{\tau_p}\right),
\end{equation}
which is reminiscent of the Walker solution for modeling the profile of 180$^\circ$ magnetic domain wall \cite{Walker1974} by replacing the time coordinate $\tau$ with the space coordinate $x$. Here $\tau_0$ is the switching time, $\tau_p$ represents the life-time of uniform magnons, and $\tau=\omega_0 t$. From perturbation theory, we derive the analytical form of these two parameters
\begin{equation}\label{taueq}
  \tau_p=-\frac{1+\alpha_\mathrm{eff}^2}{\alpha_\mathrm{eff}},~~\text{and}~~
  \tau_0=\tau_p \tanh^{-1}\sqrt{1-\frac{4B_\mathrm{opt}^2}{B_\mathrm{eff}^2}}.
\end{equation}

Figure \ref{timesolution}(c) shows the time evolution of $s_z$. Symbols represent the numerical results, dashed curves label the theoretical fittings of ansatz (\ref{tanheq}), and the solid curve is the analytical formula without fitting. The fitted switching time $\tau'_0=100.4$ ($\tau''_0=107.8$) and magnon life-time $\tau'_p=14.9$ ($\tau''_p=22.4$) from from Eq. (\ref{Seq1}) [Eq. (\ref{Seq2})] compare well with the analytical formula (\ref{taueq}) which gives $\tau_0=98.9$ and $\tau_p=22.1$. We further show that the analytical ansatz agrees excellently with numerical results in a broad range of detuning parameters, as plotted in Fig. \ref{timesolution}(d).

\begin{figure}[!htbp]
  \centering
  \includegraphics[width=\linewidth]{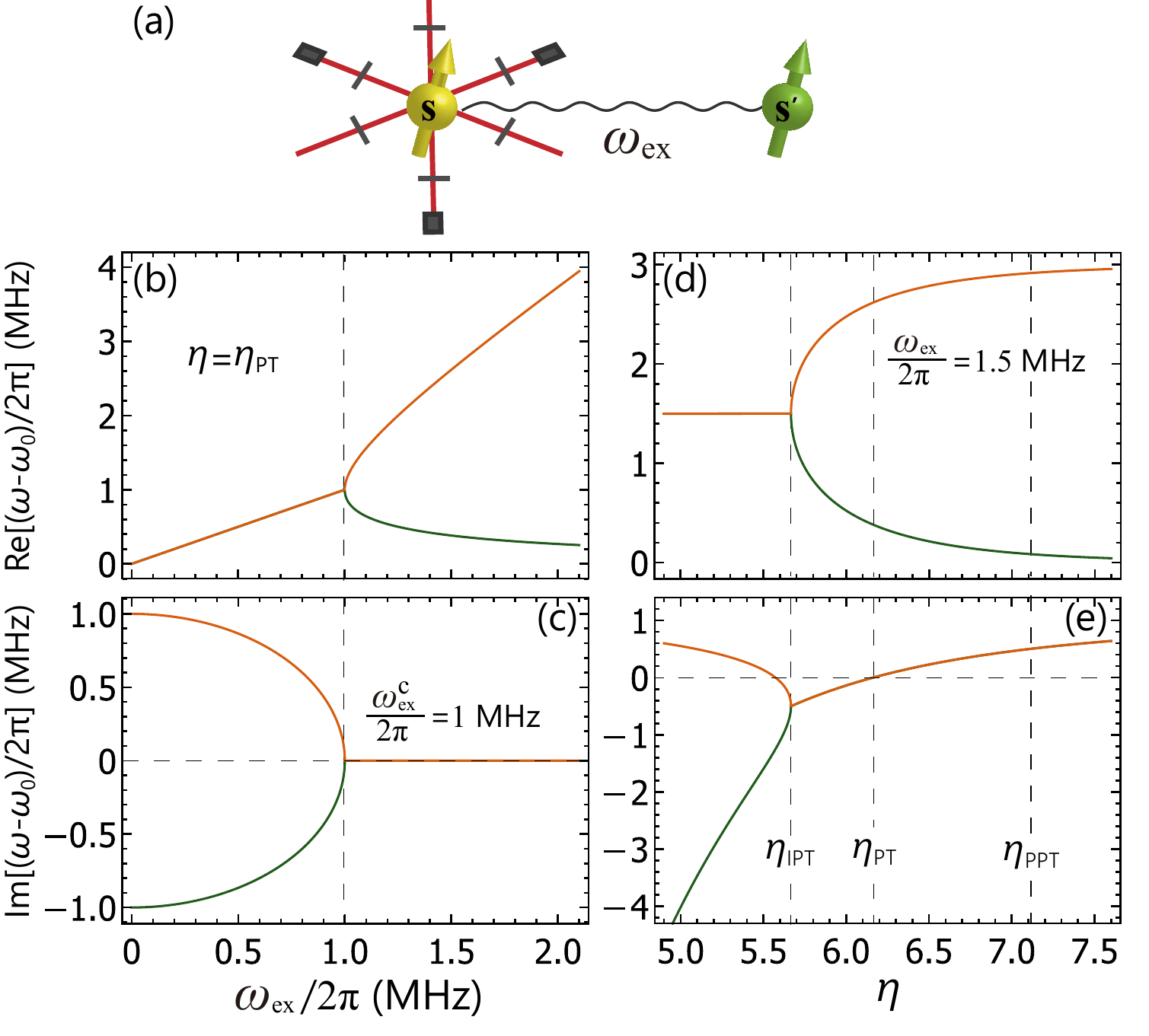}
  \caption{(a) Spin dimmer consisting of an optically pumped spin $\mathbf{s}$ and a purely lossy one $\mathbf{s}'$. Evolution of eigenfrequencies vs. the exchange coupling (b,c) at the detuning $\eta_\mathrm{PT}=6.16$, and
  vs. the detuning parameter (d,e) at the exchange coupling $\omega_\mathrm{ex}/2\pi=1.5$ MHz.}\label{ptsymmetry}
\end{figure}

\emph{Phase transition in spin dimers.}---We have shown that under proper conditions, one can realize the Gilbert-type magnetic gain which is essential for observing $\mathcal{PT}$-symmetry in purely magnetic structures. Next, we consider the optically pumped spin $\textbf{S}$ interacting with a lossy one $\textbf{S}'$, as shown in Fig. \ref{ptsymmetry}(a). The coupled spin dynamics is described by the Landau-Lifshitz-Gilbert equation
\begin{subequations}\label{pteqn}
  \begin{equation}
  \dot{\mathbf{s}}=-\gamma\mathbf{s}\times\mathbf{B}_{\text{eff}}
  +\omega_\mathrm{ex} \mathbf{s}\times\mathbf{s}'
  +\alpha_\mathrm{eff} \dot{\mathbf{s}}\times\mathbf{s},
\end{equation}
\begin{equation}
  \dot{\mathbf{s}}'=-\gamma\mathbf{s}'\times\mathbf{B}'_{\text{eff}}
  +\omega_\mathrm{ex} \mathbf{s}'\times\mathbf{s}
  +\alpha \dot{\mathbf{s}}'\times\mathbf{s}',
\end{equation}
\end{subequations}
where $\mathbf{s}^{(')}\equiv\mathbf{S}^{(')}/S$ is the unit spin vector. Since the optically induced magnetic field is the same order of magnitude with the geomagnetic field (much smaller than $B_0$), it can be safely ignored. Spin $\textbf{s}'$ is exchange coupled to the optically pumped spin $\textbf{s}$, and suffers an intrinsic Gilbert damping. If $\alpha_\mathrm{eff}=-\alpha$, the two-spin system satisfies the $\mathcal{PT}$-symmetry: Eqs. (\ref{pteqn}) are invariant in the combined operation of the parity $\mathcal{P}$ ($\mathbf{s}\leftrightarrow \mathbf{s}'$ and $\mathbf{B}_{\text{eff}}\leftrightarrow\mathbf{B}'_{\text{eff}}$) and the time reversal $\mathcal{T}$ ($t \rightarrow -t$, $\textbf{s}\rightarrow -\textbf{s}$, $\textbf{s}'\rightarrow -\textbf{s}'$, $\mathbf{B}_{\text{eff}}\rightarrow -\mathbf{B}_{\text{eff}}$, and $\mathbf{B}'_{\text{eff}}\rightarrow-\mathbf{B}'_{\text{eff}}$).

Assuming a harmonic time-dependence for the small-angle spin precession $s_{x,y}(t)=s_{x,y} e^{i\omega t}$ with $|s_{x,y}|\ll1$, one can solve the eigenspectrum of Eqs. (\ref{pteqn}). By tuning the spin-spin coupling strength $\omega_\mathrm{ex}$, we observe a transition from exact $\mathcal{PT}$ phase to the broken $\mathcal{PT}$ phase, separated by the EP at $\omega^{\text{c}}_\mathrm{ex}/2\pi=1$ MHz for $\eta=\eta_\text{\tiny PT}=6.16$, as shown in Figs. \ref{ptsymmetry}(b) and \ref{ptsymmetry}(c). Interestingly, the unequal gain and loss, i.e., $\alpha_\mathrm{eff}<0$ and $\alpha_\mathrm{eff}\neq-\alpha$, leads to an imbalanced parity-time ($\mathcal{IPT}$)-symmetry. In this region ($\eta>\eta_\mathrm{IPT}=5.66$), the eigenfrequencies have different real parts but share the identical imaginary one, as plotted in Fig. \ref{ptsymmetry}(d). A passive parity-time ($\mathcal{PPT}$)-symmetry is further identified when $\alpha_\mathrm{eff}>0$. In such case ($\eta>\eta_\mathrm{PPT}=7.11$), the imaginary part of both branches is smaller than their intrinsic damping [see Fig. \ref{ptsymmetry}(e)].

\emph{Discussion.}---In the above derivation, we focus on the case that the intrinsic Gilbert damping is isotropic. Our approach can also be generalized to treat the case when the intrinsic damping is anisotrpic \cite{Safonov2002,Chen2018b}. The three propagating lasers then should be accordingly adjusted to match the tensor form of the intrinsic magnetic damping, by modulating the driving power or the frequency of each beam, for instance. The red-detuning region is appealing to cool magnons to the subtle quantum domain. Inspired by $\mathcal{PT}$-symmetric optics \cite{Jing2014}, we envision a giant enhancement of the magnonic gain and an ultralow-threshold magnon lasing in a two-cavity system with balanced optical gain and loss, which is an open question for future study. While the magnonic passive $\mathcal{PT}$ symmetry has been observed by Liu \emph{et al}. \cite{Liu2019}, the exact and imbalanced $\mathcal{PT}$ phases are still waiting for the experimental discovery.

\emph{Conclusion.}---To summarize, we have proposed an optomagnonic method to generate the negative Gilbert damping in ferromagnets, by studying the parametric dynamics of a macrospin coupled with three orthogonally propagating circularly-polarized lasers in an optical cavity. We analytically derived the formula of the optical torque on the spin and identified the condition for the magnetic gain exactly in the Gilbert form. We found a hyperbolic-tangent function reminiscent of the Walker ansatz to well describe the time-resolved spin switching when the intrinsic damping is overcome. We finally investigated the spectrum of exchange coupled spin dimers. By varying the detuning parameter, we observed phase transitions from imbalanced to exact, and to passive $\mathcal{PT}$ symmetries. Our findings suggest an experimentally feasible way to achieve negative Gilbert damping that is essential for studying the $\mathcal{PT}$ physics and for observing the EP in magnonic systems.

\begin{acknowledgments}
We acknowledge the helpful discussion with Z. Z. Sun. This work was supported by the National Natural Science Foundation of China (Grants No. 11704060 and No. 11604041) and the National Key Research Development Program under Contract No. 2016YFA0300801.
\end{acknowledgments}

\end{document}